\documentclass[twocolumn,secnumarabic,amssymb, nobibnotes, aps, pra]{revtex4-1}

\usepackage{amsmath}
\usepackage{amsfonts}
\usepackage{amssymb}
\usepackage{graphicx}
\usepackage{braket}
\usepackage{tikz}
\usepackage{textgreek}

\usepackage{graphicx}% Include figure files
\usepackage{dcolumn}% Align table columns on decimal point
\usepackage{bm}% bold math

\begin{document}

\title{A simple approach to the quantum theory of nonlinear fiber optics}% Force line breaks with \\

\author{J. Bonetti}
\author{A. D. S\'anchez}
\author{S. M. Hernandez}
\author{D. F. Grosz}
\affiliation{Instituto Balseiro, Av. Bustillo km 9.500, Bariloche (R8402AGP), Argentina}
\affiliation{Consejo Nacional de Investigaciones Cient\'ificas y T\'ecnicas (CONICET), Argentina}

\begin{abstract}
We put forth an approach to obtain a quantum master equation for the propagation of light in nonlinear fiber optics by relying on simple quantum pictures of the processes (linear and nonlinear) occurring along propagation in an optical fiber.  This equation is shown to be in excellent agreement with the classical Generalized Nonlinear Schr\"{o}dinger Equation and predicts the effects of self-steepening and spontaneous Raman scattering. Last, we apply these results to the analysis of two cases of relevance in quantum technologies: single-photon frequency translation and spontaneous four-wave mixing. 
\end{abstract}
\maketitle

The nonlinear fiber optics realm is ruled by the Generalized Nonlinear Schr\"{o}dinger Equation (GNLSE)~\cite{agrawal2007nonlinear}, which describes the propagation of classical light pulses. This equation is not useful in quantum-technology applications since it cannot be applied to non-classical light~\cite{Li2005optical,mcguinness2010quantum,Kaiser2016fully,PengXiang2012heralded}. For these cases, a master equation describing the propagation of the quantum state of light ($\rho$), instead of the classical field, is required
\begin{equation}
\frac{\partial\rho}{\partial z} = \mathcal{F} (\rho),
\label{mystery}
\end{equation}
an equation that, to the best of our knowledge, has not yet been developed. A starting point could be the quantum theories for nonlinear fibers~\cite{Boyd2008nonlinear,Blow1991exact,Boivin1994analytical,Lai1995general,Carter1995quantum,Drummond2001quantum}, namely
\begin{equation}
\frac{\partial\rho}{\partial t} = \mathrm{Tr}_{\mathrm{fiber}}\lbrace   \frac{1}{i\hbar} [\hat{H},\bm{\rho}] \rbrace,
\label{qmechanics}
\end{equation}
where $\bm{\rho}$ is the density matrix of the system, including both light and optical fiber, $\rho = \mathrm{Tr}_{\mathrm{fiber}}\{\bm{\rho}\}$ is the reduced density matrix for the quantum state of light, in which we are interested, and  $\hat{H} = \hat{H}_{\mathrm{light}} + \hat{H}_{\mathrm{fiber}} + \hat{H}_{\mathrm{int}}$~\cite{Carter1995quantum,Drummond2001quantum}. However, this equation is not easily solvable due to the complexity of the operators involved and the large dimension of $\bm{\rho}$.  

In this paper we propose a simple approach to Eq.~(\ref{mystery}), avoiding the difficult derivation from Eq.~(\ref{qmechanics}). Our proposal relies on simple quantum pictures of the different processes that occur while propagation in a nonlinear fiber in terms of creation and annihilation of photons~\cite{Boyd2008nonlinear}. Such an approach is shown not only to be in excellent agreement with the GNLSE in the classical limit but, also, to predict the effect of self-steepening~\cite{agrawal2007nonlinear}, an aspect that has not been studied from a quantum point of view in the literature, and the effect of spontaneous Raman scattering (SpRS). Moreover, this master equation is readily applicable to relevant schemes found in quantum technologies. As an example, we apply it for a novel analysis of frequency translation of single-photons by Bragg scattering (BS)~\cite{mcguinness2010quantum} and spontaneous four-wave mixing (SpFWM), a crucial process in heralded single-photon sources~\cite{PengXiang2012heralded}.

 In the remainder of the text we use $\int$ to represent $\int_{-\infty}^{\infty}$. We start by the GNLSE describing the propagation of the complex envelope of the electrical field $A(z,t)$ in a polarization-preserving fiber  
\begin{equation}
i\frac{\partial A}{\partial z} = \left(-i\frac{\hat{\alpha}}{2}+\hat{\beta} \right)A-\gamma A \int_{0}^{\infty}R(t') \left|A(t-t') \right|^2 dt',
\label{GNLSEt}
\end{equation}
where $|A|^2$ is the optical power, $\hat{\alpha}$ and $\hat{\beta}$ are linear operators defined as $\hat{\alpha} e^{-iwt} = \alpha_w e^{-iwt}$, $\hat{\beta} e^{-iwt} = -\beta_w e^{-iwt}$. $\alpha_w$ and $\beta_w$ are the frequency profiles of loss and dispersion, respectively. $\gamma$ is the fiber nonlinear coefficient and $R(t)$ is the nonlinear-response function including both the instantaneous (electronic) and delayed Raman response~\cite{agrawal2007nonlinear}. We replace  $A(z,t) = (2\pi)^{-1/2}\int A_w(z) e^{-iwt} dw$ into Eq.~(\ref{GNLSEt}) to obtain the frequency domain version of this equation,
 \begin{multline}
\frac{\partial A_w}{\partial z}  = \left(-\frac{\alpha_w}{2}+i\beta_w\right) A_w +\\
i \tilde{\gamma}\iint R_{\mu}A^*_{w'}A_{w-\mu}A_{w'+\mu} dw' d\mu,
\end{multline}
where $\tilde{\gamma} = \gamma / 2 \pi$ and $R_w = \int R(t')e^{iwt'}dt'$.

In a similar way to the work done in Ref.~\cite{Lai1988quantumsoliton} we perform the quantization of $A$ (see Appendix A) by proposing the correspondence with the quantum operators $\hat{A}_w$, related to the annihilation and creation operators by
\begin{equation}
\hat{A}_w = \sqrt{\frac{\hbar \omega}{\Delta w}}\hat{a}_w,
\end{equation}
where $\omega = \omega_0 + w$ is the photon frequency of the envelope mode $w$, $\omega_0$ is the envelope central frequency, $\Delta w = 2\pi/T$ and $T$ is the quantization period. The number operators  $\hat{n}_{w}=\hat{a}^\dagger_{w}\hat{a}_w$ are interpreted as the quantity of photons of frequency $\omega$ within $T$.

The master equation proposed in this work is a standard Lindbland equation~\cite{Pearle2012} that reads

\begin{multline}
\frac{\partial \rho}{\partial z} = i[\hat{H}_{\mathrm{disp}}+\hat{H}_{\mathrm{fwm}},\rho]+\\
\int\hat{L}_{\textnormal{\textalpha}w'} \rho \hat{L}^{\dagger}_{\textnormal{\textalpha}w'} -\frac{1}{2} \{ \rho , \hat{L}^{\dagger}_{\textnormal{\textalpha}w'} \hat{L}_{\textnormal{\textalpha}w'}  \}dw'+\\
\int_{0}^{\infty}\hat{L}_{\mathrm{R}\mu} \rho \hat{L}^{\dagger}_{\mathrm{R}\mu} -\frac{1}{2} \{ \rho , \hat{L}^{\dagger}_{\mathrm{R}\mu} \hat{L}_{\mathrm{R}\mu}  \}d\mu,
\label{master}
\end{multline}
where
\begin{equation}
\hat{H}_{\mathrm{disp}} =  \int\frac{\beta_{w'}}{\hbar \omega'} \hat{A}^{\dagger}_{w'}\hat{A}_{w'}  dw',
\end{equation}
\begin{equation}
\hat{H}_{\mathrm{fwm}} =  \iiint \frac{\tilde{\gamma}R^R_\mu}{2\hbar \omega_0}\hat{A}^{\dagger}_{w_1}\hat{A}^{\dagger}_{w_2}\hat{A}_{w_1-\mu}\hat{A}_{w_2+\mu} dw_1 dw_2 d\mu,
\end{equation}
\begin{equation}
\hat{L}_{\textnormal{\textalpha} w'} = \sqrt{\frac{ \alpha_{w'}}{\hbar \omega'}} \hat{A}_{w'},
\end{equation}
\begin{equation}
\hat{L}_{\mathrm{R}\mu} = \int\sqrt{\frac{2\tilde{\gamma} R^I_\mu}{\hbar \omega_0}}\hat{A}^{\dagger}_{w'-\mu}\hat{A}_{w'} dw'.
\end{equation}
$R^R_\mu$ and $R^I_\mu$ are respectively the real and imaginary parts of $R_\mu$. In this equation we can easily recognize the four basic processes of the GNLSE: loss ($\hat{L}_{\textnormal{\textalpha}}$), dispersion ($\hat{H}_{\mathrm{disp}}$), four-wave mixing ($\hat{H}_{\mathrm{fwm}}$) and stimulated Raman scattering (SRS) ($\hat{L}_{\mathrm{R}}$).

We test this master equation by calculating the evolution of the mean values $\braket{\hat{A}_w}$. For this it is important the commutation relations of the operators $\hat{A}_w$, that in the limit $T \rightarrow \infty$ read $[\hat{A}_w,\hat{A}_{w'}] = 0$ and $[\hat{A}_w,\hat{A}^\dagger_{w'}] = \hbar\omega \delta(w-w')$. The calculation is quite involved (see Appendix B) but straightforward and gives

\begin{multline}
\frac{\partial}{\partial z} \braket{\hat{A}_w} = \left(-\frac{\alpha_w}{2}+i\beta_w\right) \braket{ \hat{A}_w }+\\
i \tilde{\gamma} \left(1+\frac{w}{\omega_0}\right)\iint R_{\mu}\braket{\hat{A}^\dagger_{w'}\hat{A}_{w-\mu}\hat{A}_{w'+\mu} }dw' d\mu-\\
\tilde{\gamma} \left(1+\frac{w}{\omega_0}\right)\int_{0}^{\infty} R^I_\mu \hbar (\omega-\mu) d\mu\braket{\hat{A}_{w}}.
\label{meanvalues}
\end{multline}

Observe that the evolution of the mean values is in excellent agreement with the GNLSE. Further, the factor $\left(1+w/\omega_0\right)$, that introduces a frequency dependence of the nonlinear coefficient $\tilde{\gamma}$, does not represent a departure from the classical limit. Instead, it appears naturally in the derivation and accounts for the well-known phenomenon in nonlinear fiber optics of \textit{self-steepening}~\cite{agrawal2007nonlinear}. 
Although often associated with the distortion of ultra-short pulses, recent work has shown this effect to be of importance in the context of modulation instability and its complex interplay with stimulated Raman scattering~\cite{Sanchez2018tunable,Sanchez2018anti}. It must be emphasized that the effect of self-steepening has been overlooked in earlier quantum theories underlying the GNLSE, but it is \textit{built in} in Eq.~(\ref{master}).   

The operator $\hat{L}_{\mathrm{R}}$ in the master equation gives a clear picture of the Raman scattering process, interpreted as the annihilation of a photon at the pump frequency $w'$ and the creation of a photon at the signal frequency $w'-\mu$. This process can occur even when there are no photons present at the signal, producing the effect known as spontaneous Raman scattering (SpRS). It is the source of the last term in the Eq.~(\ref{meanvalues}) and cannot be obtained from the classical GNLSE. We use this term to calculate the loss of a pump, at frequency $\omega_0$, induced by SpRS and obtain 
\begin{equation}
\frac{\partial I_p}{\partial z} = -2\tilde{\gamma} \left[\int_{0}^{\infty} R^I_\mu \hbar (\omega_0-\mu) d \mu\right] I_p.
\end{equation}
This result is in agreement with earlier works on Raman scattering~\cite{Friis2017effects,Rottwitt2003scaling} .

Despite the large size of the density matrix, we can use the master equation to study quantum-technology devices. In these applications we are interested in few-photons cases and hence the density matrix is reduced to numerically manageable sizes. We analyze two relevant cases depicting the interaction of few-photons signals with strong pumps. The quantum propagation of these small signals is derived from Eq.~(\ref{master}) by considering the pumps as classical and undepleted, and regarding only frequencies of interest.  

\begin{figure}
\centering
\begin{tikzpicture}
\draw[thick,->] (7.5,0) -- (12.5,0);
\node at (12.2,-0.2) (jugh) {$w$};
\draw[thick,->] (10,0) -- (10,3);
\node at (9.7,2.7) (fhnei){$p$};
\draw[thick,->,blue] (8.5,0) -- (8.5,1.5);
\node at (8.7,1.2) (fhnei){$i$};
\draw[thick,->,green!50!black] (11.5,0) -- (11.5,1.5);
\node at (11.7,1.2) (fhnei){$s$};
\draw[thick,<->,dashed,gray] (10,0.3) -- (11.5,0.3);
\node at (8.6,-0.15) (hfn){\small{$w_s$}};
\node at (11.6,-0.15) (hfn){\small{$w_i$}};
\node at (10.1,-0.15) (hfn){\small{$w_p$}};
\node at (10.8,0.5) (hfn){\small{$\Omega$}};
\draw[thick,<->,dashed,gray] (8.5,0.3) -- (10,0.3);
\node at (9.3,0.5) (hfn){\small{$\Omega$}};
\end{tikzpicture}
\caption{Frequency configuration of a degenerate four-wave mixing process.}
\label{fig3}
\end{figure}
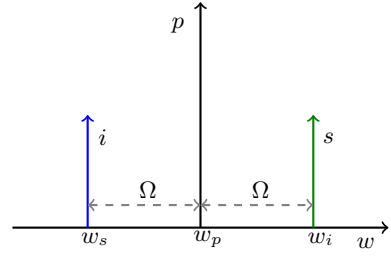

\begin{figure}
\begin{center}
\includegraphics[width=\columnwidth]{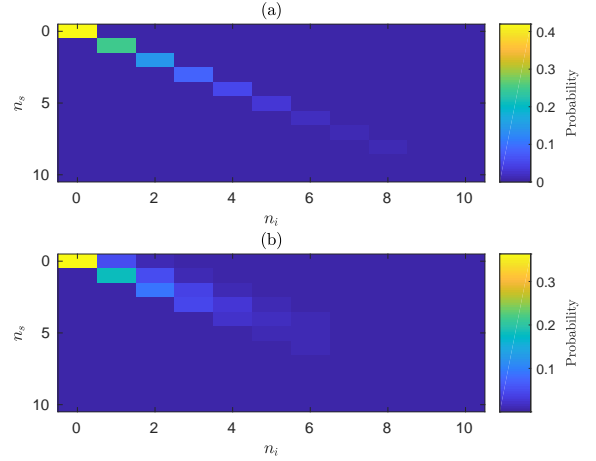}
\end{center}
\caption{Degenerate SpFWM without (a) and with (b) loss and SRS. Probabilities of off-diagonal elements in (b) seriously affect the heralded-photon scheme.}
\label{figheralded}
\end{figure}

 Degenerate SpFWM, showed in Fig.~\ref{fig3} is a parametric non-classical phenomenon by which two photons of a noiseless CW pump ($p$) are annihilated and simultaneously two photons are created in the signal ($s$) and the idler ($i$). This effect can be exploited to design heralded-photon sources by detecting one of the created photons to infer the presence of the other~\cite{PengXiang2012heralded}. Loss and SRS affect this scheme, producing false heralding or non-heralded photons.

We simplify the master equation by proposing $\hat{A}_w = \sqrt{2\pi P}e^{i k_{p}z}\delta(w-w_p) + \hat{A}_{w_s}\delta_{w w_s} + \hat{A}_{w_i}\delta_{w w_i}$, $\alpha_{w_p} = 0$ and the phase-matching condition to obtain (see Appendix C)
\begin{equation}
\frac{\partial \rho}{\partial z} = i[\hat{H},\rho] + \sum_{\nu=1}^{3} \hat{L}_\nu \rho \hat{L}^{\dagger}_\nu-\frac{1}{2}\{\rho , \hat{L}^{\dagger}_\nu\hat{L}_\nu \},
\end{equation}
where
\begin{equation}
\hat{H} = k_s \hat{b}^\dagger_s\hat{b}_s+k_i \hat{b}^\dagger_i\hat{b}_i+ \gamma P R^R_\Omega \left(\hat{b}^{\dagger}_s \hat{b}^\dagger_i + \hat{b}_s\hat{b}_i\right),
\end{equation}
\begin{equation}
\hat{L}_1 = \sqrt{\alpha_s} \hat{b}_{s},
\end{equation}
\begin{equation}
\hat{L}_2 = \sqrt{\alpha_i} \hat{b}_{i},
\end{equation}
\begin{equation}
\hat{L}_3 = \sqrt{2\gamma P R^I_\Omega}\left(\hat{b}_s + \hat{b}^\dagger_i\right).
\end{equation}
In this case $k_p = \beta_{w_p} + \gamma P$, $k_x = \beta_{w_x} + \gamma P (1+R^R_\Omega)$, and phase-matching means $2k_p-k_s-k_i = 0$.

In Fig.~\ref{figheralded} we show numerical results for a particular case. The parameters are those of the frequency translation example. The probability of measuring $n_s$ photons in the frequency $s$ and $n_i$ photons in the frequency $i$ is calculated by $\braket{n_s n_i|\rho |n_s n_i}$. Loss and SRS seriously affect the heralding scheme by the possibility of mismatch between the photon numbers $n_s$ and $n_i$. This simple calculation proves useful to accurately evaluate the performance of a heralded-photon scheme in the presence of loss and SRS.

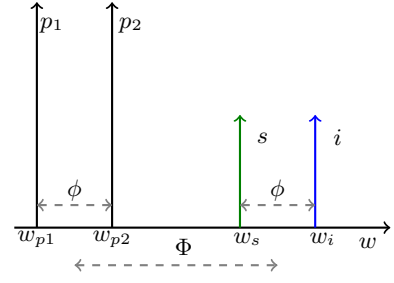
\begin{figure}
\centering
\begin{tikzpicture}
\draw[thick,->] (7,0) -- (12,0);
\node at (11.7,-0.2) (jugh) {$w$};
\draw[thick,->] (7.3,0) -- (7.3,3);
\node at (7.5,2.7) (fhnei){$p_1$};
\draw[thick,->] (8.3,0) -- (8.3,3);
\node at (8.55,2.7) (fhnei){$p_2$};
\draw[thick,->,green!50!black] (10,0) -- (10,1.5);
\node at (10.3,1.2) (fhnei){$s$};
\draw[thick,->,blue] (11,0) -- (11,1.5);
\node at (11.3,1.2) (fhnei){$i$};
\draw[thick,<->,dashed,gray] (7.8,-0.5) -- (10.5,-0.5);
\draw[thick,<->,dashed,gray] (10,0.3) -- (11,0.3);
\node at (9.25,-0.25) (hfn){\small{$\Phi$}};
\node at (7.3,-0.15) (hfn){\small{$w_{p1}$}};
\node at (8.3,-0.15) (hfn){\small{$w_{p2}$}};
\node at (10.1,-0.15) (hfn){\small{$w_s$}};
\node at (11.1,-0.15) (hfn){\small{$w_i$}};
\node at (10.5,0.5) (hfn){\small{$\phi$}};
\draw[thick,<->,dashed,gray] (7.3,0.3) -- (8.3,0.3);
\node at (7.8,0.5) (hfn){\small{$\phi$}};
\end{tikzpicture}
\caption{Frequency configuration of the Bragg scattering process.}
\label{fig1}
\end{figure}

 Bragg Scattering (BS), showed in Fig.~\ref{fig1} is a well-known four-wave mixing process by which two strong pumps ($p_1$ and $p_2$) interact with two small signals (signal $s$ and idler $i$). The quantum mechanical description of BS implies the annihilation of one photon in $s$ and the simultaneous creation of one photon in $i$. Particularly, single-photon inputs in BS processes are of significant importance in quantum technologies~\cite{mcguinness2010quantum}. To the best of our knowledge, the analysis of single-photon translation including loss and SRS has not been reported in the literature before due to the difficulty in including these effects in quantum models~\cite{Friis2017effects}.
 
As in the previous case, we simplify Eq.~(\ref{master}) by proposing $\hat{A}_w = \sqrt{2\pi P}e^{i k_{p1}z}\delta(w-w_{p1})+ \sqrt{2\pi P}e^{i k_{p2}z}\delta(w-w_{p2}) + \hat{A}_{w_s}\delta_{w w_s} + \hat{A}_{w_i}\delta_{w w_i}$,
where $P$ is the optical power of the pumps, and $k_{px}=\beta_{w_x}+\gamma P(2+R^R_\phi)$. We assume that the pumps do not interact by SRS ($R^I_{\phi}=0$) and  $\alpha_{w_{p1}} = \alpha_{w_{p2}} = 0$. The simplified master equation for this particular case reads
\begin{equation}
\frac{\partial \rho}{\partial z} = i[\hat{H},\rho] + \sum_{\nu=1}^{5} \hat{L}_\nu \rho \hat{L}^{\dagger}_\nu-\frac{1}{2}\{\rho , \hat{L}^{\dagger}_\nu\hat{L}_\nu \},
\end{equation}
where
\begin{multline}
\hat{H} = k_s \hat{b}^\dagger_s\hat{b}_s+k_i \hat{b}^\dagger_i\hat{b}_i+\\
\gamma P\left(R^R_\phi+ R^R_\Phi \right)\left(\hat{b}^{\dagger}_s \hat{b}_i + \hat{b}^{\dagger}_i \hat{b}_s\right),
\end{multline}
\begin{equation}
\hat{L}_1 = \sqrt{\alpha_s} \hat{b}_{s},
\end{equation}
\begin{equation}
\hat{L}_2 = \sqrt{\alpha_i} \hat{b}_{i},
\end{equation}
\begin{equation}
\hat{L}_3 = \sqrt{2\gamma P R^I_{\Phi-\phi} } \hat{b}_s,
\end{equation}
\begin{equation}
\hat{L}_4 = \sqrt{2\gamma P R^I_{\Phi+\phi} } \hat{b}_i,
\end{equation}
\begin{equation}
\hat{L}_5 = \sqrt{2\gamma P R^I_\Phi}\left(\hat{b}_s + \hat{b}_i\right).
\end{equation}
We introduced the annihilation operators $\hat{b}_x = e^{-i k_x z}\hat{a}_{w_x}$, with $k_s = \beta_{w_s} + 2\gamma P(R^R_{\Phi}+R^R_{\Phi-\phi})$, $k_i = \beta_{w_i} + 2\gamma P(R^R_{\Phi}+R^R_{\Phi+\phi})$ and assumed the perfect phase-matching condition $k_{p1}-k_{p2}+k_i-k_s = 0$.

\begin{figure}
\begin{center}
\includegraphics[width=\columnwidth]{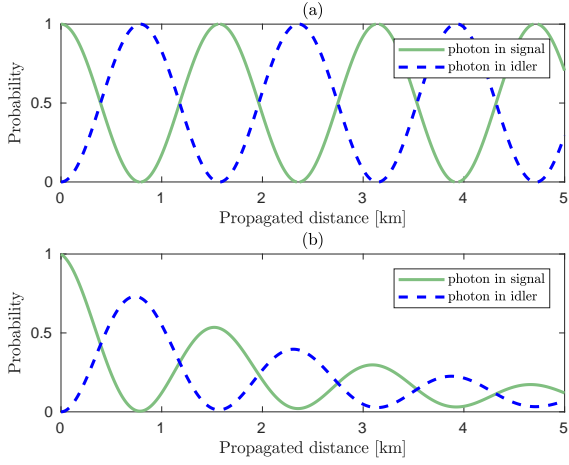}
\end{center}
\caption{Propagation of a single-photon in a frequency-translation scheme, without (a) and with (b) loss and SRS.}
\label{fig2}
\end{figure}

We numerically solve this system for the single-photon input with $\gamma = 1$ W$^{-1}$km$^{-1}$, $P=1$ W, $\alpha_{w_s} = \alpha_{w_i} =  0.01$ km$^{-1}$, $R^R_\Phi = R^R_\phi = 1$, $R^I_\Phi = R^I_{\Phi-\phi} = R^I_{\Phi+\phi} = 0.1$, $\omega_s \simeq \omega_i \simeq \omega_0$ and a fiber length of $5$ km. In Fig. \ref{fig2} we show the probabilities of finding the photon at $w_s$ or $w_i$, calculated by $\braket{10|\rho|10}$ and $\braket{01|\rho|01}$. This calculation can be used to evaluate the efficiency of the single-photon frequency translation process even when loss and SRS effects are taken into account.

In conclusion, we proposed a simple approach to a quantum theory of light propagation in nonlinear fiber optics. By computing the evolution of the field-operator mean values we showed excellent agreement with the classical GNLSE. Further, the proposed quantum master equation leads naturally to the effects of spontaneous Raman scattering and self-steepening, the latter being included for the first time in a quantum framework. Finally, we applied the proposed approach to two cases of singular relevance in quantum technologies: single-photon frequency translation and spontaneous four-wave mixing.

\begin{widetext}
\begin{footnotesize}
\section*{Appendix A}
\label{app:A}
A quantization of the field $A$ is carried out in a similar fashion as standard quantum optics texts, i. e., by establishing an analogue between optical modes and harmonic oscillators. Thus, we propose $A(t)$ to be a periodic solution of period $T$ that can be represented by the (Fourier) series
\begin{equation*}
A(t) = \sum_{m=-\infty}^{\infty}O_m(t)e^{i\omega_0 t},
\end{equation*}
where
\begin{equation*}
O_m(t) = \frac{\Delta w}{\sqrt{2\pi}}A_m e^{-i\omega_m t},
\end{equation*}
with $\Delta w = 2\pi/T$, $\omega_m =\omega_0 + m \Delta w$,  and $\omega_0$ is the central frequency of the envelope $A$. We consider the $O_m(t)$ as solutions to oscillators 
\begin{equation*}
\frac{\partial}{\partial t} O_m = -i\omega_m O_m.
\end{equation*}
The energy carried by each oscillator within the period $T$ is $E_m = T |O_m|^2$. By decomposing these solutions as $O_m = O^R_m+iO^I_m$, we obtain equations
\begin{equation}
\frac{\partial}{\partial t}O^R_m = \omega_m O^I_m, \qquad \frac{\partial}{\partial t}O^I_m = -\omega_m O^R_m, \qquad E_m = T\left((O^R_m)^2+(O^I_m)^2\right).
\label{oscillator}
\end{equation}
Introducing variables $x_m = \sqrt{\frac{2T}{M\omega_m^2}}O^R_m$ and $p_m = \sqrt{2MT}O^I_m$, where $M$ is an arbitrary mass, and replacing them into Eqs.~(\ref{oscillator}) we obtain a classical harmonic oscillator
\begin{equation*}
\frac{\partial x_m}{\partial t} = \frac{p_m}{M}, \qquad \frac{\partial p_m}{\partial t} = -M\omega_m^2 x_m, \qquad E_m = \frac{M \omega_m^2}{2}x_m^2+\frac{p_m^2}{2M},
\end{equation*}
whose quantization is given by the operators $\hat{H}_m$, $\hat{x}_m$ and $ \hat{p}_m$ such that 
\begin{equation*}
\hat{H}_m = \frac{M\omega_m^2}{2}\hat{x}_m^2+\frac{\hat{p}_m^2}{2M}, \qquad [\hat{x}_m,\hat{p}_m]=i\hbar.
\end{equation*}
An annihilation operator is defined correspondingly as
\begin{equation}
\label{eq:ann1}
\hat{a'}_m = \sqrt{\frac{M \omega_m}{2\hbar}}\left(\hat{x}_m + \frac{i}{M \omega_m}\hat{p}_m \right),
\end{equation}
and the number operator $\hat{n}_m = \hat{a'}^\dagger_m \hat{a'}_m$ is interpreted as the quantity of photons of frequency $\omega_m$ within the period $T$.

From Eq.~(\ref{eq:ann1}) we can obtain operators $\hat{A}_m$ by noting that
$$ \hat{a'}_m = \sqrt{\frac{T}{\hbar \omega_m}}\hat{O}_m = \sqrt{\frac{\Delta w}{\hbar \omega_m}}\hat{A}_m e^{-i\omega_mt},
$$
and also deduce the commutation relations
\begin{equation*}
[\hat{A}_m,\hat{A}_n] = [\hat{A}^\dagger_m,\hat{A}^\dagger_n] = 0,
\end{equation*}
\begin{equation*}
[\hat{A}_m,\hat{A}^\dagger_n] = \frac{\hbar \omega_m}{\Delta w} \delta_{nm},
\end{equation*}
where $\delta_{nm}$ is the Kroenecker's delta. In the limit where $T \rightarrow \infty$, $\Delta w \rightarrow 0$ we can deduce that $[\hat{A}_\omega,\hat{A}_{\omega'}^\dagger] = \hbar \omega \delta(\omega-\omega')$.

When dealing with number states, it is useful to define annihilation operators
\begin{equation*}
\hat{a}_m = \hat{a'}_m e^{i \omega_m t} = \sqrt{\frac{\Delta w}{\hbar \omega_m}}\hat{A}_m,
\end{equation*}
for as it can be readily noticed,  $\hat{n}_m = \hat{a'}^\dagger_m\hat{a'}_m = \hat{a}^\dagger_m\hat{a}_m$.
\end{footnotesize}

\begin{footnotesize}
\section*{Appendix B}
In the following we derive the evolution of the mean value of $\hat{A}_w$. For the sake of clarity, we deal separately with the four quantum processes that add up to the right hand side of Eq.~\ref{master}.

\subsection*{Dispersion}

\begin{equation*}
i[\hat{A}_w,\hat{H}_{\textnormal{\tiny{disp}}}] = i\left[ \hat{A}_w ,  \int \frac{\beta_{w'}}{\hbar \omega'} \hat{A}^{\dagger}_{w'}\hat{A}_{w'} \partial w'\right] = \\
i\int\frac{\beta_{w'}}{\hbar \omega'}\left[ \hat{A}_w ,    \hat{A}^{\dagger}_{w'}\right]\hat{A}_{w'} \partial w' = i\int\beta_{w'}\delta (w-w')\hat{A}_{w'} \partial w' = i\beta_{w}\hat{A}_{w}.
\end{equation*}
\\[5 pt]
\begin{equation*}
\frac{\partial}{\partial z} \braket{A_w} = \braket{i[\hat{A}_w,\hat{H}_{\textnormal{\tiny{disp}}}]} = i\beta_{w} \braket{\hat{A}_{w}}.
\end{equation*}

\subsection*{Four-wave mixing}

\begin{equation*}
i[\hat{A}_w,\hat{H}_{\textnormal{\tiny{FWM}}}] = i\left[ \hat{A}_w ,  \tilde{\gamma} \iiint\frac{R^R_\mu}{2\hbar \omega_0}\hat{A}^{\dagger}_{w_1}\hat{A}^{\dagger}_{w_2}\hat{A}_{w_1-\mu}\hat{A}_{w_2+\mu} \partial w_1 \partial w_2 \partial \mu\right] =\\ 
i \tilde{\gamma} \iiint \frac{R^R_\mu}{2\hbar \omega_0}\left[ \hat{A}_w , \hat{A}^{\dagger}_{w_1}\hat{A}^{\dagger}_{w_2}\hat{A}_{w_1-\mu}\hat{A}_{w_2+\mu} \right]\partial w_1 \partial w_2 \partial \mu = 
\end{equation*}
\begin{equation*}
i \tilde{\gamma} \iiint \frac{R^R_\mu}{2\hbar \omega_0}\left(\left[ \hat{A}_w , \hat{A}^{\dagger}_{w_1} \right]\hat{A}^{\dagger}_{w_2}\hat{A}_{w_1-\mu}\hat{A}_{w_2+\mu}+\hat{A}^{\dagger}_{w_1}\left[ \hat{A}_w , \hat{A}^{\dagger}_{w_2} \right]\hat{A}_{w_1-\mu}\hat{A}_{w_2+\mu}\right)\partial w_1 \partial w_2 \partial \mu = 
\end{equation*}
\begin{equation*}
i \tilde{\gamma} \iiint \frac{R^R_\mu \omega}{2 \omega_0}\left(\delta(w-w_1)\hat{A}^{\dagger}_{w_2}\hat{A}_{w_1-\mu}\hat{A}_{w_2+\mu}+\hat{A}^{\dagger}_{w_1}\delta(w-w_2)\hat{A}_{w_1-\mu}\hat{A}_{w_2+\mu}\right)\partial w_1 \partial w_2 \partial \mu = 
\end{equation*}
\begin{equation*}
i \tilde{\gamma} \iint \frac{R^R_\mu \omega}{2 \omega_0}\hat{A}^{\dagger}_{w_2}\hat{A}_{w-\mu}\hat{A}_{w_2+\mu} \partial w_2 \partial \mu+i \tilde{\gamma} \iint \frac{R^R_\mu \omega}{2 \omega_0}\hat{A}^{\dagger}_{w_1}\hat{A}_{w_1-\mu}\hat{A}_{w+\mu}\partial w_1  \partial \mu = 
\end{equation*}
\begin{equation*}
i \tilde{\gamma} \iint \frac{R^R_\mu \omega}{2 \omega_0}\hat{A}^{\dagger}_{w'}\hat{A}_{w-\mu}\hat{A}_{w'+\mu} \partial w' \partial \mu+i \tilde{\gamma} \iint \frac{R^R_\mu \omega}{2 \omega_0}\hat{A}^{\dagger}_{w'}\hat{A}_{w+\mu}\hat{A}_{w'-\mu}\partial w'  \partial \mu = 
\end{equation*}
\begin{equation*}
i \tilde{\gamma} \iint \frac{R^R_\mu \omega}{2 \omega_0}\hat{A}^{\dagger}_{w'}\hat{A}_{w-\mu}\hat{A}_{w'+\mu} \partial w' \partial \mu+i \tilde{\gamma} \iint \frac{R^R_{\nu} \omega}{2 \omega_0}\hat{A}^{\dagger}_{w'}\hat{A}_{w-\nu}\hat{A}_{w'+\nu}\partial w'  \partial \nu =\\ i \tilde{\gamma}\frac{\omega}{\omega_0} \iint R^R_\mu \hat{A}^{\dagger}_{w'}\hat{A}_{w-\mu}\hat{A}_{w'+\mu} \partial w' \partial \mu.
\end{equation*}
where, in the second to the last line, we used the change of variable $\nu = -\mu$ in the second integral and, also, we used the Fourier transform property $R^R_{-\mu}=R^R_{\mu}$.
\\[5 pt]

\begin{equation*}
\frac{\partial}{\partial z} \braket{A_w} = \braket{i[\hat{A}_w,\hat{H}_{\textnormal{\tiny{FWM}}}]}= i \tilde{\gamma}\frac{\omega}{\omega_0} \iint R^R_\mu \braket{\hat{A}^{\dagger}_{w'}\hat{A}_{w-\mu}\hat{A}_{w'+\mu}} \partial w' \partial \mu.
\end{equation*}

\subsection*{Losses}
\begin{equation*}
\left[\hat{L}^\dagger_{\alpha w'},\hat{A}_w \right]\hat{L}_{\alpha w'} = \left[\sqrt{\frac{ \alpha_{w'}}{\hbar \omega'}} \hat{A}^\dagger_{w'},\hat{A}_w \right]\sqrt{\frac{ \alpha_{w'}}{\hbar \omega'}} \hat{A}_{w'} =\\ \frac{2 \alpha_{w'}}{\hbar \omega'} \left[ \hat{A}^\dagger_{w'},\hat{A}_w \right] \hat{A}_{w'} = -2\alpha_{w'}\delta (w-w')\hat{A}_{w'}.
\end{equation*}
\\[5pt]

\begin{equation*}
\hat{L}^\dagger_{\alpha w'}\left[\hat{L}_{\alpha w'},\hat{A}_w \right] =\sqrt{\frac{ \alpha_{w'}}{\hbar \omega'}} \hat{A}^\dagger_{w'} \left[\sqrt{\frac{ \alpha_{w'}}{\hbar \omega'}} \hat{A}_{w'},\hat{A}_w \right] = 0.
\end{equation*}
\\[5pt]

\begin{equation*}
\frac{\partial}{\partial z} \braket{\hat{A}_w} = \frac{1}{2}\int \braket{\left[\hat{L}^\dagger_{\alpha w'},\hat{A}_w \right]\hat{L}_{\alpha w'}}-\braket{\hat{L}^\dagger_{\alpha w'}\left[\hat{L}_{\alpha w'},\hat{A}_w \right]}\partial w' =\\ \frac{1}{2}\braket{\int - \alpha_{w'}\delta (w-w')\hat{A}_{w'}\partial w'} = -\frac{\alpha_{w}}{2}\braket{\hat{A}_{w}}.
\end{equation*}

\subsection*{Raman scattering}

\begin{equation*}
[\hat{L}^{\dagger}_{R\mu},\hat{A}_w] = \left[\int\sqrt{\frac{2\tilde{\gamma} R^I_{\mu}}{\hbar \omega_0}}\hat{A}^{\dagger}_{w'} \hat{A}_{w'-\mu}\partial w',\hat{A}_w\right] = \int\sqrt{\frac{2\tilde{\gamma} R^I_{\mu}}{\hbar \omega_0}}\left[\hat{A}^{\dagger}_{w'}\hat{A}_{w'-\mu} ,\hat{A}_w\right]\partial w' =
\end{equation*}
\begin{equation*}
- \sqrt{\frac{2 \tilde{\gamma} R^I_\mu}{\hbar \omega_0}}\int \hbar \omega \delta (w-w')\hat{A}_{w'-\mu}\partial w' = - \sqrt{\frac{2\tilde{\gamma} R^I_{\mu}}{\hbar \omega_0}}\hbar \omega \hat{A}_{w-\mu}.
\end{equation*}
\\[5pt]

\begin{equation*}
[\hat{L}^{\dagger}_{R\mu},\hat{A}_w]\hat{L}_{R\mu} =  - \sqrt{\frac{2\tilde{\gamma} R^I_{\mu}}{\hbar w_0}}\hbar \omega \hat{A}_{w-\mu}\int\sqrt{\frac{2\tilde{\gamma} R^I_{\mu}}{\hbar \omega_0}}\hat{A}^{\dagger}_{w'-\mu}\hat{A}_{w'} \partial w' =\\ - 2\tilde{\gamma} \frac{\omega}{\omega_0}R^I_{\mu} \int\hat{A}_{w-\mu}\hat{A}^{\dagger}_{w'-\mu}\hat{A}_{w'} \partial w'=
\end{equation*}
\begin{equation*}
- 2\tilde{\gamma} \frac{\omega}{\omega_0}R^I_{\mu}  \int [\hat{A}_{w-\mu},\hat{A}^{\dagger}_{w'-\mu}]\hat{A}_{w'} \partial w'- 2\tilde{\gamma} \frac{\omega}{\omega_0}R^I_{\mu}  \int\hat{A}^{\dagger}_{w'-\mu}\hat{A}_{\omega-\mu}\hat{A}_{w'} \partial w' =
\end{equation*}
\begin{equation*}
- 2\tilde{\gamma} \frac{\omega}{\omega_0}R^I_{\mu} \left( \int\hbar (\omega-\mu) \delta(w-w')\hat{A}_{w'} \partial w'+\int\hat{A}^{\dagger}_{w'-\mu}\hat{A}_{w-\mu}\hat{A}_{w'} \partial w'\right)=
\end{equation*}
\begin{equation*}
- 2\tilde{\gamma} \frac{\omega}{\omega_0}R^I_{\mu} \left( \hbar (\omega-\mu) \hat{A}_{w}+\int\hat{A}^{\dagger}_{w'}\hat{A}_{w-\mu}\hat{A}_{w'+\mu} \partial w'\right).
\end{equation*}
\\[5pt]

\begin{equation*}
\int_{0}^{\infty} [\hat{L}^{\dagger}_{R\mu},\hat{A}_w]\hat{L}_{R\mu} \partial \mu = -\int_{0}^{\infty} 2\tilde{\gamma} \frac{\omega}{\omega_0}R^I_{\mu} \left( \hbar (\omega-\mu) \hat{A}_{w}+\int\hat{A}^{\dagger}_{w'}\hat{A}_{w-\mu}\hat{A}_{w'+\mu} \partial w'\right)\partial \mu =
\end{equation*}
\begin{equation*}
-\int_{0}^{\infty}\frac{2\tilde{\gamma} R^I_\mu \hbar \omega(\omega-\mu)}{\omega_0} \hat{A}_{w} \partial \mu-\int_{0}^{\infty}\int 2\tilde{\gamma} \frac{\omega}{\omega_0}R^I_{\mu}\hat{A}^{\dagger}_{w'}\hat{A}_{w-\mu}\hat{A}_{w'+\mu} \partial w'\partial \mu.
\end{equation*}
\\[5pt]

\begin{equation*}
[\hat{L}_{R\mu},\hat{A}_w] = \left[\int\sqrt{\frac{2\tilde{\gamma} R^I_{\mu}}{\hbar \omega_0}}\hat{A}^{\dagger}_{w'-\mu}\hat{A}_{w'} \partial w',\hat{A}_w\right] = \int\sqrt{\frac{2\tilde{\gamma} R^I_{\mu}}{\hbar \omega_0}}\left[\hat{A}^{\dagger}_{w'-\mu} \hat{A}_{w'},\hat{A}_w\right]\partial w' =
\end{equation*}
\begin{equation*}
- \sqrt{\frac{2\tilde{\gamma} R^I_{\mu}}{\hbar \omega_0}}\int \hbar \omega \delta (w-w'+\mu)\hat{A}_{w'}\partial w' = - \sqrt{\frac{2\tilde{\gamma} R^I_{\mu}}{\hbar w_0}} \hbar \omega \hat{A}_{w+\mu}.
\end{equation*}
\\[5pt]

\begin{equation*}
\hat{L}^{\dagger}_{R\mu}[\hat{L}_{R\mu},\hat{A}_w] = \int\sqrt{\frac{2\tilde{\gamma} R^I_{\mu}}{\hbar \omega_0}}\hat{A}^{\dagger}_{w'}\hat{A}_{w'-\mu} \partial w'\left(- \sqrt{\frac{2\tilde{\gamma} R^I_{\mu}}{\hbar \omega_0}}\hbar \omega \hat{A}_{w+\mu}\right) =\\ -2\tilde{\gamma} \frac{\omega}{\omega_0}R^I_{\mu} \int\hat{A}^{\dagger}_{w'}\hat{A}_{w'-\mu} \hat{A}_{w+\mu}\partial w' =
-2\tilde{\gamma} \frac{\omega}{\omega_0}R^I_{\mu} \int\hat{A}^{\dagger}_{w'} \hat{A}_{w+\mu}\hat{A}_{w'-\mu}\partial w'.
\end{equation*}
\\[5pt]

\begin{equation*}
\int_{0}^{\infty}\hat{L}^{\dagger}_{R\mu}[\hat{L}_{R\mu},\hat{A}_w]\partial \mu = -\int_{0}^{\infty} 2\tilde{\gamma} \frac{\omega}{\omega_0}R^I_{\mu}\int\hat{A}^{\dagger}_{w'} \hat{A}_{w+\mu}\hat{A}_{w'-\mu}\partial w'\partial \mu = 
\end{equation*}
\begin{equation*}
\int_{0}^{-\infty}\int2\tilde{\gamma} \frac{\omega}{\omega_0}R^I_{-\nu}\hat{A}^{\dagger}_{w'} \hat{A}_{w-\nu}\hat{A}_{w'+\nu}\partial w'\partial \nu =
\end{equation*}

\begin{equation*}
\int_{-\infty}^{0}\int 2\tilde{\gamma} \frac{\omega}{\omega_0}R^I_{\mu}\hat{A}^{\dagger}_{w'} \hat{A}_{w+\mu}\hat{A}_{w'-\mu}\partial w'\partial \mu.
\end{equation*}
\\[5pt]

\begin{equation*}
\int_{0}^{\infty}[\hat{L}^{\dagger}_{R\mu},\hat{A}_w]\hat{L}_{R\mu} -\hat{L}^{\dagger}_{R\mu}[\hat{L}_{R\mu},\hat{A}_w]  \partial \mu =\\ -\iint 2\tilde{\gamma} \frac{\omega}{\omega_0}R^I_{\mu}\hat{A}^{\dagger}_{w'} \hat{A}_{w+\mu}\hat{A}_{w'-\mu}\partial w'\partial \mu -\int_{0}^{\infty}\frac{2\tilde{\gamma} R^I_\mu \hbar \omega(\omega-\mu)}{\omega_0} \hat{A}_{w} \partial \mu.
\end{equation*}
\\[5pt]

\begin{equation*}
\frac{\partial}{\partial z} \braket{\hat{A}_w} = \frac{1}{2}\int \braket{[\hat{L}^{\dagger}_{R\mu},\hat{A}_w]\hat{L}_{R\mu}} -\braket{\hat{L}^{\dagger}_{R\mu}[\hat{L}_{R\mu},\hat{A}_w] }\partial w' =\\  -\tilde{\gamma} \frac{\omega}{\omega_0}\iint R^I_{\mu}\braket{\hat{A}^{\dagger}_{w'} \hat{A}_{w+\mu}\hat{A}_{w'-\mu}}\partial w'\partial \mu -\tilde{\gamma} \frac{\omega}{\omega_0} \int_{0}^{\infty} R^I_\mu \hbar (\omega -\mu)  \partial \mu\braket{\hat{A}_{w}}.
\end{equation*}

\section*{Appendix C}

We consider a modified version of Eq.~\ref{master} for periodic signals, i.e.,
\begin{equation*}
\frac{\partial \rho}{\partial z} = i\left[\hat{H}_{\mathrm{disp}}+\hat{H}_{\mathrm{fwm}},\rho\right]+
\sum_{m}\hat{L}_{\textnormal{\textalpha}m'} \rho \hat{L}^{\dagger}_{\textnormal{\textalpha}m} -\frac{1}{2} \{ \rho , \hat{L}^{\dagger}_{\textnormal{\textalpha}m} \hat{L}_{\textnormal{\textalpha}m}  \}\Delta w+\\
\sum_{n=0}^{\infty}\hat{L}_{\mathrm{R}n} \rho \hat{L}^{\dagger}_{\mathrm{R}n} -\frac{1}{2} \{ \rho , \hat{L}^{\dagger}_{\mathrm{R}n} \hat{L}_{\mathrm{R}n}  \}\Delta w,
\end{equation*}
where
\begin{equation*}
\hat{H}_{\mathrm{disp}} =  \sum_{m}\frac{\beta_{w_m}}{\hbar \omega_m} \hat{A}^{\dagger}_{m}\hat{A}_{m}  \Delta w,
\end{equation*}
\begin{equation*}
\hat{H}_{\mathrm{fwm}} =  \sum_{m,n,j} \frac{\tilde{\gamma}R^R_{w_j}}{2\hbar \omega_0}\hat{A}^{\dagger}_{m}\hat{A}^{\dagger}_{n}\hat{A}_{m-j}\hat{A}_{n+j} \Delta w^3,
\end{equation*}
\begin{equation*}
\hat{L}_{\textnormal{\textalpha} m} = \sqrt{\frac{ \alpha_{w_m}}{\hbar \omega_m}} \hat{A}_{m},
\end{equation*}
\begin{equation*}
\hat{L}_{\mathrm{R}n} = \sum_{m}\sqrt{\frac{2\tilde{\gamma} R^I_{w_n}}{\hbar \omega_0}}\hat{A}^{\dagger}_{m-n}\hat{A}_{m} \Delta w.
\end{equation*}
We propose the operator
\begin{equation*}
    \hat{A}_m = \frac{\sqrt{2\pi P}}{\Delta w}e^{i k_{p}z}\delta_{m0} + \hat{A}_{s}\delta_{ms} + \hat{A}_{i}\delta_{mi}.
\end{equation*}
The dispersion hamiltonian reads
\begin{equation*}
\hat{H}_{\mathrm{disp}} =  \sum_{m}\frac{\beta_{w_m}}{\hbar \omega_m} \left(\frac{\sqrt{2\pi P}}{\Delta w}e^{-i k_{p}z}\delta_{m0} + \hat{A}_{s}^\dagger \delta_{ms} + \hat{A}_{i}^\dagger\delta_{mi} \right)\\
\left(\frac{\sqrt{2\pi P}}{\Delta w}e^{i k_{p}z}\delta_{m0} + \hat{A}_{s}\delta_{ms} + \hat{A}_{i}\delta_{mi}\right) \Delta w,
\end{equation*}
\begin{equation} \label{eq:hdisp_per}
\hat{H}_{\mathrm{disp}} =  \frac{2\pi P\beta_{w_0}}{\hbar \omega_0} + \frac{\beta_{w_s}\hat{A}_s^\dagger\hat{A}_s}{\hbar \omega_s}\Delta w + \frac{\beta_{w_i}\hat{A}_i^\dagger\hat{A}_i}{\hbar \omega_i}\Delta w.
\end{equation}
The first term can be neglected since it is a constant and does not affect the motion equations for $\hat{A}_s$ and $\hat{A}_i$. Equation \ref{eq:hdisp_per} can be expressed more succintly as
\begin{equation*}
\hat{H}_{\mathrm{disp}} =  \beta_{w_s}\hat{a}_s^\dagger\hat{a}_s + \beta_{w_i}\hat{a}_i^\dagger\hat{a}_i.
\end{equation*}

The FWM hamiltonian reads
\begin{multline*}
\hat{H}_{\mathrm{fwm}} =  \sum_{m,n,j} \frac{\tilde{\gamma}R^R_{w_j}}{2\hbar \omega_0}\left(A_0^*\delta_{m0} + \hat{A}_{s}^\dagger \delta_{ms} + \hat{A}_{i}^\dagger \delta_{mi}\right)\\ \left( A_0^*\delta_{n0} + \hat{A}_{s}^\dagger \delta_{ns} + \hat{A}_{i}^\dagger \delta_{ni}\right)\left( A_0\delta_{(m-j)0} + \hat{A}_{s}\delta_{(m-j)s} + \hat{A}_{i}\delta_{(m-j)i}\right)\\
\left(A_0\delta_{(n+j)0} + \hat{A}_{s}\delta_{(n+j)s} + \hat{A}_{i}\delta_{(n+j)i}\right) \Delta w^3,
\end{multline*}
where
\begin{equation*}
    A_0 = \frac{\sqrt{2\pi P}}{\Delta w}e^{ik_pz}.
\end{equation*}
Just as done with the dispersion hamiltonian, we neglect the constant term. Also, we neglect terms with products of three to four operators $\hat{A}_s$ or $\hat{A}_i$, since signal or idler optical powers are negligible against pump's powers. Hence, using the relation $R^R_{-w} = R^R_{w}$ we obtain
\begin{equation*}
\hat{H}_{\mathrm{fwm}} =  \frac{\tilde{\gamma}}{2\hbar \omega_0}\left(2(1+R^R_{w_s})|A_0|^2(\hat{A}_s^\dagger\hat{A}_s + \hat{A}_i^\dagger\hat{A}_i) + 2R^R_{w_s}(A_0^2\hat{A}_s^\dagger \hat{A}_i^\dagger+(A_0^*)^2\hat{A}_s \hat{A}_i)\right)\Delta w^3,
\end{equation*}
\begin{equation*}
\hat{H}_{\mathrm{fwm}} =  \frac{\gamma P}{\hbar \omega_0}\left((1+R^R_{w_s})(\hat{A}_s^\dagger\hat{A}_s + \hat{A}_i^\dagger\hat{A}_i) + R^R_{w_s}(e^{2ik_pz}\hat{A}_s^\dagger \hat{A}_i^\dagger+e^{-2ik_pz}\hat{A}_s \hat{A}_i)\right)\Delta w.
\end{equation*}
Replacing with creation and annihilation operators, and assuming that the frequencies  $\omega_s$ and $\omega_i$ are very close to $\omega_0$, we obtain
\begin{equation*}
\hat{H}_{\mathrm{fwm}} =  \gamma P\left((1+R^R_{w_s})(\hat{a}_s^\dagger\hat{a}_s + \hat{a}_i^\dagger\hat{a}_i) + R^R_{w_s}(e^{2ik_pz}\hat{a}_s^\dagger \hat{a}_i^\dagger+e^{-2ik_pz}\hat{a}_s \hat{a}_i)\right).
\end{equation*}
We introduce new creation-annihilation operators, defined as
\begin{equation*}
     \hat{b}_s = e^{-i k_s z}\hat{a}_{s},
\end{equation*}
\begin{equation*}
     \hat{b}_i = e^{-i k_i z}\hat{a}_{i},
\end{equation*}
where
\begin{equation*}
    k_s = \beta_{w_s} + \gamma P (1+R^R_{w_s}),
\end{equation*}
\begin{equation*}
    k_i = \beta_{w_i} + \gamma P (1+R^R_{w_i}).
\end{equation*}
Having in mind the phase-matching condition, $2k_p-k_s-k_i = 0$, the hamiltonian reads
\begin{equation*}
\hat{H}_{\mathrm{fwm}} =  \gamma P\left((1+R^R_{w_s})(\hat{b}_s^\dagger\hat{b}_s + \hat{b}_i^\dagger\hat{b}_i) + R^R_{w_s}(\hat{b}_s^\dagger \hat{b}_i^\dagger+\hat{b}_s \hat{b}_i)\right).
\end{equation*}
This two hamiltonians can be compacted into a unique operator,
\begin{equation*}
    \hat{H} = \hat{H}_{\mathrm{disp}} + \hat{H}_{\mathrm{fwm}} = k_s\hat{b}_s^\dagger\hat{b}_s + k_i\hat{b}_i^\dagger\hat{b}_i + \gamma PR^R_{w_s}(\hat{b}_s^\dagger \hat{b}_i^\dagger+\hat{b}_s \hat{b}_i). 
\end{equation*}

Attenuation operators, only relevant for $s$-and-$i$-indexed modes, read
\begin{equation*}
    \hat{L}_{\textnormal{\textalpha}s} = \sqrt{\frac{\alpha_{w_s}}{\hbar\omega_s}}\hat{A}_s.
\end{equation*}
\begin{equation*}
    \hat{L}_{\textnormal{\textalpha}i} = \sqrt{\frac{\alpha_{w_i}}{\hbar\omega_i}}\hat{A}_i,
\end{equation*}
We use the property
\begin{equation*}
    \left(\hat{L}_{\textnormal{\textalpha}m} \rho \hat{L}^{\dagger}_{\textnormal{\textalpha}m} -\frac{1}{2} \{ \rho , \hat{L}^{\dagger}_{\textnormal{\textalpha}m} \hat{L}_{\textnormal{\textalpha}m}  \}\right)\Delta w = \hat{L'}_{\textnormal{\textalpha}m} \rho \hat{L'}^{\dagger}_{\textnormal{\textalpha}m} -\frac{1}{2} \{ \rho , \hat{L'}^{\dagger}_{\textnormal{\textalpha}m} \hat{L'}_{\textnormal{\textalpha}m} \},
\end{equation*}
where
\begin{equation*}
    \hat{L'}_{\textnormal{\textalpha}m} = \sqrt{\alpha_{w_m}}\hat{b}_m.
\end{equation*}

Raman-scattering operators read
\begin{equation*}
\hat{L}_{\mathrm{R}n} = \sum_{m}\sqrt{\frac{2\tilde{\gamma} R^I_{w_n}}{\hbar \omega_0}}\left(A_0^*\delta_{(m-n)0} + \hat{A}_{s}^\dagger\delta_{(m-n)s} + \hat{A}_{i}^\dagger\delta_{(m-n)i} \right)\left(A_0\delta_{m0} + \hat{A}_{s}\delta_{ms} + \hat{A}_{i}\delta_{mi}\right) \Delta w.
\end{equation*}
We neglect operators with $\hat{A}^2_s$, $\hat{A}^2_i$ or $\hat{A}_s\hat{A}_i$, then we only keep
\begin{equation*}
\hat{L}_{\mathrm{R}s} = \sum_{m}\sqrt{\frac{2\tilde{\gamma} R^I_{w_s}}{\hbar \omega_0}}\left(A_0^*\delta_{(m-s)0} + \hat{A}_{s}^\dagger\delta_{(m-s)s} + \hat{A}_{i}^\dagger\delta_{(m-s)i} \right)\left(A_0\delta_{m0} + \hat{A}_{s}\delta_{ms} + \hat{A}_{i}\delta_{mi}\right) \Delta w,
\end{equation*}
\begin{equation*}
\hat{L}_{\mathrm{R}s} = \sqrt{\frac{2\tilde{\gamma} R^I_{w_s}}{\hbar \omega_0}}\left(A_0^*\hat{A}_s + A_0 \hat{A}_i^\dagger \right) \Delta w,
\end{equation*}
\begin{equation*}
\hat{L}_{\mathrm{R}s} = \sqrt{\frac{2\gamma P R^I_{w_s}}{\hbar \omega_0}}\left(e^{-ik_pz}\hat{A}_s +  e^{ik_pz}\hat{A}_i^\dagger \right) ,
\end{equation*}

It can be easily shown that
\begin{equation*}
    \left(\hat{L}_{\mathrm{R}s} \rho \hat{L}^{\dagger}_{\mathrm{R}s} -\frac{1}{2} \{ \rho , \hat{L}^{\dagger}_{\mathrm{R}s} \hat{L}_{\mathrm{R}s}  \}\right)\Delta w = \hat{L'}_{\mathrm{R}s} \rho \hat{L'}^{\dagger}_{\mathrm{R}s} -\frac{1}{2} \{ \rho , \hat{L'}^{\dagger}_{\mathrm{R}s} \hat{L'}_{\mathrm{R}s}  \},
\end{equation*}
where
\begin{equation*}
\hat{L'}_{\mathrm{R}s} = \sqrt{2\gamma P R^I_{w_s}}\left(\hat{b}_s + \hat{b}_i^\dagger \right).
\end{equation*}
We assumed that $\omega_s \simeq \omega_i \simeq \omega_0$.

Finally, the master equation for this special case is
\begin{equation*}
\frac{\partial \rho}{\partial z} = i[\hat{H},\rho] + \sum_{\nu=1}^{3} \hat{L}_\nu \rho \hat{L}^{\dagger}_\nu-\frac{1}{2}\{\rho , \hat{L}^{\dagger}_\nu\hat{L}_\nu \},
\end{equation*}
with
\begin{equation*}
\hat{H} = k_s \hat{b}^\dagger_s\hat{b}_s+k_i \hat{b}^\dagger_i\hat{b}_i+ \gamma P R^R_{w_s}  \left(\hat{b}^{\dagger}_s \hat{b}^\dagger_i + \hat{b}_s\hat{b}_i\right),
\end{equation*}
\begin{equation*}
\hat{L}_1 = \sqrt{\alpha_s} \hat{b}_{s},
\end{equation*}
\begin{equation*}
\hat{L}_2 = \sqrt{\alpha_i} \hat{b}_{i},
\end{equation*}
\begin{equation*}
\hat{L}_3 = \sqrt{2\gamma P R^I_{w_s}}\left(\hat{b}_s + \hat{b}^\dagger_i\right).
\end{equation*}
\end{footnotesize}

\end{widetext}

\end{document}